\documentclass[twocolumn,prl,showpacs,amsmath,amssymb]{revtex4}
\usepackage{graphicx}

\begin{document}

\title{Resonant emergence of global and local spatiotemporal order in a nonlinear field model}

\author{Marcelo Gleiser}
\email{mgleiser@dartmouth.edu}

\author{Rafael C. Howell}
\email{rhowell@dartmouth.edu}

\affiliation{Department of Physics and Astronomy, Dartmouth College,
Hanover, NH 03755, USA}

\date{\today}

\begin{abstract}
We investigate the nonequilibrium evolution of a scalar field in (2+1) dimensions. The field is set in
a double-well potential in contact (open) or not (closed) with a heat bath. For closed systems, we
observe the synchronized emergence of coherent spatiotemporal configurations, identified with
oscillons. This initial global ordering degenerates into localized order until all oscillons disappear.
We show that the synchronization is driven by resonant parametric oscillations of the field's zero mode
and that local ordering is only possible outside equipartition. None of these orderings occur for open
systems.
\end{abstract}


\maketitle

The emergence of spatiotemporal ordered structures in nonlinear systems is an ideal laboratory for
investigating the trend toward complexification observed in nature at the physical, chemical, and
biological level \cite{walgraef}. For these ordered structures to survive, they must interact with
an external environment, which maintains the local nonequilibrium conditions. Examples can be found
in hydrodynamics, in networks of chemical reactions \cite{cross}, and in living organisms
\cite{schnerb}.

In field theory and cosmology, most of the interest in ordered configurations has focused on
topological or nontopological static solutions of the equations of motion \cite{rajaraman}. An
exception to this trend are oscillons, long-lived time-dependent localized field configurations that
have been found in field theory \cite{gleiser,bettinson}, soft condensed-matter systems 
\cite{umbanhowar}, and stellar interiors \cite{umurhan}. Here, we show how oscillons spontaneously 
emerge as a nonlinear scalar field approaches thermal equilibrium. We observe not only the local 
emergence of spatiotemporal order (oscillons), but also that this emergence is initially synchronized 
(global). Our results are applicable to any system modeled by a scalar order parameter $\phi$ with 
amplitude-dependent nonlinearities, if its potential $V(\phi)$ satisfies $\partial^2V/\partial\phi^2<0$ 
for a range of $\phi$.

Consider a (2+1) dimensional real scalar field $\phi({\bf x},t)$, which evolves under the equation of 
motion
\begin{equation}
\frac{\partial^2\phi}{\partial t^2}-\nabla^2\phi= -\frac{\partial V_{\rm dw}}{\partial\phi}.
\label{diffeq}
\end{equation}
\noindent
The double-well potential $V_{\rm dw}(\phi)=\frac{1}{4}(\phi^2-1)^2$ has minima at $\phi=\pm 1$. It has 
been shown that this system can generate oscillons, characterized by a persistent oscillatory behavior 
at their core \cite{gleiser}. To see this, prepare the field with a Gaussian profile, 
$\phi(r) = \phi_a \exp(-r^2/R^2) - 1$, and let it evolve via Eq. (\ref{diffeq}). All that is needed is 
that $V_{\rm dw}^{''}(\phi_a)<0$ and that the initial radius $R$ be larger than a bifurcation value 
$R_{\rm osc}$ \cite{gleiser,sornborger}. The properties of these so-called deterministic oscillons have 
been extensively studied in two \cite{sornborger,adib} and three dimensions \cite{gleiser,honda}.

We investigate open (canonical) and closed (microcanonical) systems. In both cases, the field is 
initially thermalized in a single-well potential $V_{\rm sw}(\phi) = (\phi+1)^2$, symmetric about
$\phi=-1$, chosen so that $V_{\rm sw}^{''}(-1) = V_{\rm dw}^{''}(-1)$. The thermalization is achieved 
by coupling the field to an external heat bath via a Langevin equation
\begin{equation}
\frac{\partial^2 \phi}{\partial t^2} + \gamma \frac{\partial \phi}{\partial t} - \nabla^2 \phi
= -V_{\rm sw}^{'}(\phi) + \xi,
\label{lang}
\end{equation}
\noindent
where the viscosity coefficient $\gamma$ is related to the stochastic force of zero mean 
$\xi({\bf x},t)$ by the fluctuation-dissipation relation ($k_B=1$ and $T$ is the temperature of the 
heat bath), 
$\langle \xi({\bf x},t)\xi({\bf x'},t') \rangle = 2 \gamma T \delta^2({\bf x}-{\bf x'}) \delta(t-t')$.
The numerical evolution is implemented on a square lattice with periodic boundary conditions, using a 
staggered leapfrog method. Integrations are performed with $\delta x = 0.1$ and $\delta t = 0.01$ with 
1024 lattice sites per side. The coupling to the heat bath continues until equipartition is satisfied 
at temperature $T$. Results are ensemble averages from 50 independent realizations, which, given the large
number of degrees of freedom, have energy $E$ within $\Delta E/E<10^{-3}$.

The potential is then switched from $V_{\rm sw}$ to $V_{\rm dw}$, and the system is tossed again out 
of equilibrium. This switch can be interpreted as an instantaneous quench on the system, implemented 
by varying some control parameter $\alpha$, as is customarily done in many applications. To see this,
shift $\phi \rightarrow \phi+1$ so that $V_{\rm sw}=\phi^2$ and 
$V_{\rm dw}=1/4[\phi^2(\alpha\phi-2)^2|_{\alpha=1}$. Note that when $\alpha=0$, 
$V_{\rm dw} = V_{\rm sw}$. The quench then occurs instantaneously by sending $\alpha=0$ to $\alpha=1$ 
(or equivalently, $V_{\rm sw} \rightarrow V_{\rm dw}$).

For the closed system the coupling to the bath is removed at the time of the switch (by setting
$\gamma=0$), while for the open system it is kept on throughout the simulation. After the switch occurs,
energy exchange between the nonlinearly coupled modes will again drive the system to equipartition.
(We note that the initial and final equilibrium temperatures never differ more than $\sim 0.15\% $. 
Thus, each experiment is referred to by its initial thermalization temperature.) For open systems, 
equilibrium is attained within a time scale of ${\cal O}(\gamma^{-1})$, without any emergence of 
ordered configurations. This is in marked contrast with closed systems.

\begin{figure}
\includegraphics[width=245pt,height=220pt]{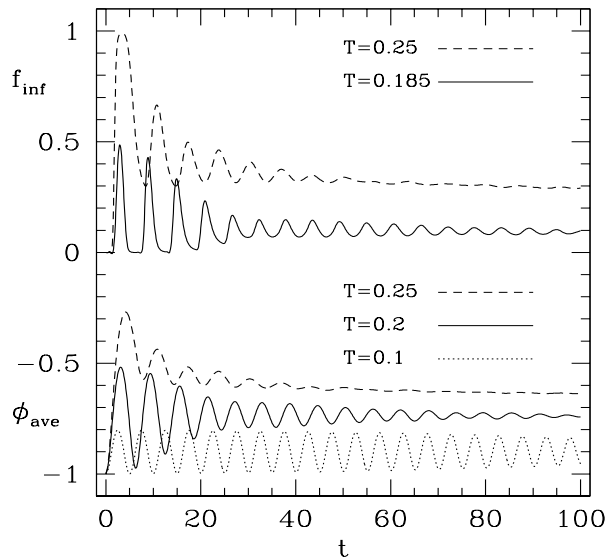}
\caption{ \label{ave}
Bottom: lattice-averaged field for various temperatures.
Top: fraction of the field above the inflection point.
}
\end{figure}

Figure \ref{ave} (lower half) shows the area-averaged field $\phi_{\rm av}(t)$ for closed systems at
various initialization temperatures. The time is set so that the switch to the double-well potential occurs at $t=0$. Note 
that $\phi_{\rm av}(t)$ displays damped oscillations, indicating that the zero mode of the field 
transfers its energy to higher modes until it reaches its temperature-dependent final equilibrium value 
$\langle\phi\rangle_T$. At low temperatures, such as $T=0.1$, the oscillations have small amplitude 
and remain nearly periodic. At higher temperatures, such as $T=0.2$, $\phi_{\rm av}(t)$ oscillates 
beyond the left inflection point and fluctuations in the field probe the unstable region of the 
double-well potential. This can also be seen in the upper half of Fig. \ref{ave}, where the time 
evolution of the fraction of the field above the inflection point $f_{\rm inf}(t)$ is shown. For 
temperatures above $T\simeq 0.185$, over half the field probes the unstable region. For even larger 
temperatures, $T\gtrsim 0.25$, the whole field goes above the inflection point, signaling the approach 
to criticality. Above this temperature the field separates into large, slowly evolving thin-walled 
domains. (The critical temperature of this system, where $\phi_{\rm av}\rightarrow 0$, is 
$T_c \simeq 0.270\pm 0.005$.)

\begin{figure}
\includegraphics[width=245pt,height=220pt]{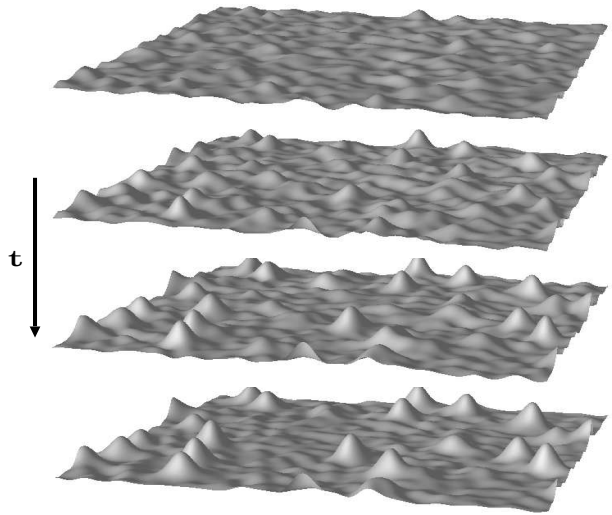}
\caption{ \label{snap}
Snapshots displaying synchronous emergence of oscillons in the 2-dimensional field, starting at
$t=15.5$ and spanning half the period of an oscillon. Simulations can be viewed in Ref. \cite{web}.
}
\end{figure}

Figure \ref{snap} shows a sequence of snapshots of the field at $T=0.2$, spanning in time about half an 
oscillation period of an oscillon. In order to generate this figure and relate the emergent
configurations to oscillons, the field is smoothed with an optimal (Wiener) filter
\cite{numericalrecipes}. This filtering technique is especially useful in our situation, since its 
implementation  requires knowledge of the unwanted thermal noise in the system. With our choice of 
thermal initial conditions, Boltzmann statistics provides us with the power spectrum at the beginning 
of the experiment, $\langle |\phi({\bf k},t=0)|^2 \rangle = T/(k^2+2)$, which is to be removed by the 
filter. Throughout the experiment, only modes with $0< |{\bf k}| \lesssim 0.8$ amplify above the noise, 
sometimes by as much as two orders of magnitude. Their spectrum has a shape and width that coincide 
with those of deterministic oscillons. The transformation back into real space clearly reveals the 
emergence of localized field configurations within a smooth background.

With the filtered field, we can catalog and track the location of all local extrema at each instant in 
time, the vast majority of which are seen to correspond to the centers of localized configurations. We 
then measure the value of the field at each extrema and the corresponding configuration radius (at 
half maximum). With ample sorting, we compile a library containing each large-amplitude fluctuation 
during the entire evolution of the system ($0<t<1500$), from which we can obtain their nucleation 
times, sizes, periods of oscillation, and lifetimes. A deterministic oscillon is characterized both by 
its large-amplitude oscillations (the field at its core probes the positive half of the potential, 
$\phi_a>1$) and its extreme longevity. We thus establish two criteria to select the subset of all 
configurations which correspond to oscillons: they must have a maximum amplitude above the background 
satisfying $\phi_a>1$ and they must survive for at least ten oscillations ($t>60$).

\begin{figure}
\includegraphics[width=245pt,height=220pt]{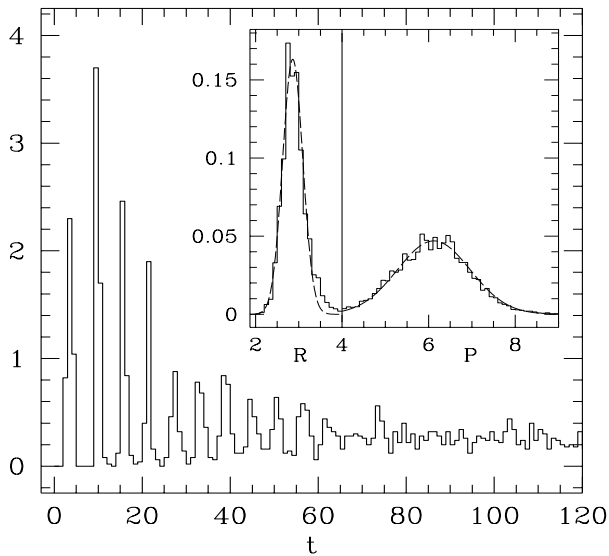}
\caption{ \label{nuc}
The number of oscillons nucleated between $t$ and $t+\delta t$ at $T=0.2$, with $\delta t=1$.
The global emergence is evident early in the simulations. Inset: the
probability distribution of radii and periods of oscillation.
}
\end{figure}

The inset in Fig. \ref{nuc} shows the probability distribution functions for the radii $R$ and periods 
of oscillation $P$ for all the oscillons present throughout the simulations, with binwidth 
$\delta R=\delta P=0.1$. The fitted curves are Gaussian functions with centers $R_0=2.86$ and 
$P_0=6.12$ and widths $\sigma_R=0.33$ and $\sigma_P=1.19$, respectively.

In Fig. \ref{nuc} we also show the distribution of nucleation times for these oscillons at $T=0.2$.
This function gives the number of oscillons nucleated between $t$ and $t+\delta t$, with $\delta t=1$. 
The sharp peaks at early times, $t<60$, correspond to the synchronous emergence of oscillons, while 
for $t>60$ this global ordering gives way to only local ordering, in which oscillons emerge at 
arbitrary times with similar probability. Even this local emergence disappears after approximately 
$t \gtrsim 500$ (discussed below). Notice the correlation between the nucleation activity of oscillons 
and the energy loss from the zero mode (Fig. \ref{ave}) at early times. This is observed for all 
temperatures within $0.185\lesssim T \lesssim 0.25$.

\begin{figure}
\includegraphics[width=245pt,height=220pt]{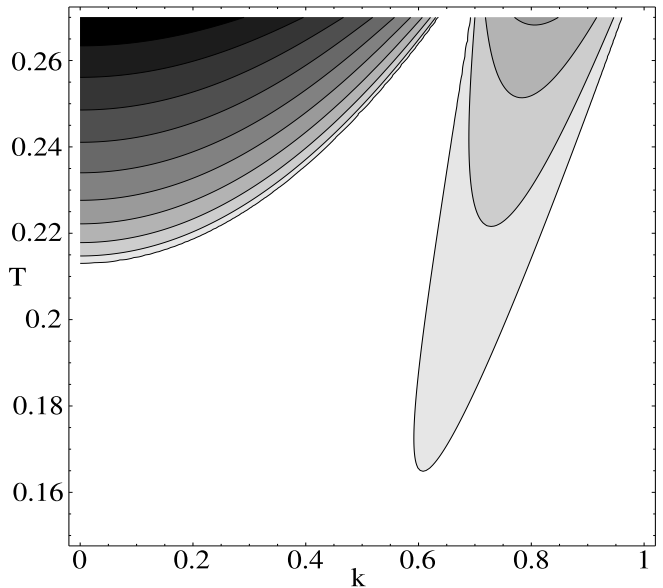}
\caption{ \label{parametric}
Lines of constant amplification rate for small-amplitude modes at various temperatures, beginning with 
$\eta^{-1}\simeq 27$ for the bottom-most contour and increasing in increments of $\delta\eta=0.05$.
}
\end{figure}

To understand the origin of this synchronous emergence, we decompose the field as
$\phi({\bf x},t)=\phi_{\rm av}(t)+\delta\phi({\bf x},t)$ and first investigate the behavior of
$\phi_{\rm av}(t)$, prior to relinquishing its energy to higher modes, by approximating
$\delta\phi({\bf x},t)$ as remaining statistically equivalent to its initial symmetric state. Upon
substituting this form of $\phi({\bf x},t)$ into Eq. (\ref{diffeq}) and performing a statistical average 
over the fluctuations, we arrive at the mean-field equation of motion for $\phi_{\rm av}(t)$:
$\ddot{\phi}_{\rm av} = [1-3 \langle\delta\phi^2\rangle]\phi_{\rm av} - \phi_{\rm av}^3$, where 
$\langle\delta\phi^2\rangle$ is the two-point correlation function at $t=0$ and depends linearly on the 
temperature $T$. Thus, $\phi_{\rm av}(t)$ starts at $\phi=-1$ and oscillates
anharmonically in the left-hand well of an effective double-well potential, with minima at
$\phi_{\pm}=\pm\sqrt{1-3 \langle\delta\phi^2\rangle}$. This mean-field approximation works very well
at describing the evolution of $\phi_{\rm av}(t)$ at temperatures sufficiently far from $T_c$ (c.f. 
$T=0.1$ in Fig. \ref{ave}). It also gives $\langle\phi\rangle_T \approx \phi_{-}$ to within $4\%$ even 
at the highest temperatures we considered.

We now examine the behavior of small fluctuations about $\phi_{\rm av}(t)$. Linearizing Eq.
(\ref{diffeq}) with respect to $\delta\phi({\bf x},t)$ and taking the Fourier transform we obtain
(for $k>0$)
\begin{equation}
\label{lineareq}
\ddot{\delta\phi}(k,t)+\left [k^2+V_{\rm dw}^{''}\left (\phi_{\rm av}(t)\right )\right ]~\delta\phi(k,t)= 0~.
\end{equation}
Equations of this type, generalized Mathieu equations, are known to exhibit parametric resonance, 
which can lead to exponential amplification ($\sim\exp{\eta t}$) in the oscillations of
$\delta\phi(k,t)$ at certain wavelengths, in response to the time-dependent harmonic term. These 
equations have been of great interest in reheating studies of inflationary cosmologies \cite{linde}. 
To verify that this is the mechanism behind the synchronous amplification of oscillon modes early in 
the simulations, we use in Eq. (\ref{lineareq}) the anharmonic solution for $\phi_{\rm av}(t)$ obtained 
in the mean-field approximation above.

Figure \ref{parametric} shows lines of constant amplification rate $\eta$ of the fluctuations 
$\delta\phi(k,t)$ for various $k$ and $T$. At low temperatures, $T<0.165$, no modes are 
ever amplified.  As the temperature is increased, so is the amplitude of oscillation in 
$\phi_{\rm av}(t)$, eventually causing the band $0.6 \lesssim k \lesssim 0.9$ to resonate. We 
note that the characteristic wavenumber of oscillon configurations is $k_{\rm osc} = 2/R_0$ 
\cite{gleiser}. Using the result in Fig. \ref{nuc} for the average configuration radius $R_0=2.86$, 
we obtain $k_{\rm osc} \simeq 0.7$. From Fig. \ref{parametric}, modes with $k_{\rm osc}\simeq 0.7$
are excited for $T\gtrsim 0.18$, the temperatures above which we see the synchronized emergence of 
oscillons. For $T>0.22$, a second band of longer wavelength modes becomes excited as well, signaling 
the onset of criticality. These general results corroborate our earlier findings: for temperatures 
$0.185\lesssim T \lesssim 0.25$, oscillations of the zero mode drive, via parametric resonance, 
amplification of the modes comprising oscillon configurations, ultimately leading to the usual 
breakdown of the linear approximation. The growth of these modes corresponds to large-amplitude 
fluctuations that probe the unstable regions of the potential ($\partial^2V/\partial\phi^2<0$), which 
coordinate to form the observed coherent structures.

\begin{figure}
\includegraphics[width=245pt,height=220pt]{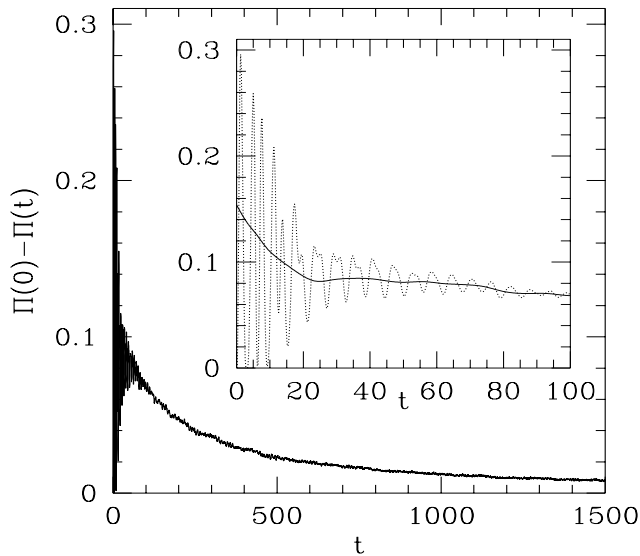}
\caption{ \label{S}
The change of $\Pi(t)$ from the initial state for closed systems at $T=0.2$. The
exponential approach to equilibrium is clear at late times. The inset illustrates the role of oscillons 
as a bottleneck to equipartition.
}
\end{figure}

Finally, we introduce a measure of the partitioning of the kinetic energy $\Pi(t)$, which we use to describe the
nonequilibrium evolution of the system:
\begin{equation}
\Pi(t)=-\int d^2 k~p({\bf k},t) \ln p({\bf k},t),
\end{equation}
where $p({\bf k},t)=K({\bf k},t)/\int d^2 k K({\bf k},t)$ and 
$K({\bf k},t)$ is the kinetic energy of the $k$th mode. $\Pi(t)$ attains its 
maximum [$\Pi_{\rm max}=\ln(N)$ on a lattice with $N$ degrees of freedom] when equipartition is satisfied.
This occurs both at the initial 
thermalization $(t=0)$ and final equilibrium states, since in this case all modes carry the same fractional 
kinetic energy. In Fig. \ref{S} we show the change of $\Pi(t)$ from the initial state, 
$\Pi(t=0)-\Pi(t)$, for the closed system at $T=0.2$. At late times ($t \gtrsim 150$), we have found that
the system equilibrates exponentially in a time scale $\tau \simeq 500$. At early times, the localization of 
energy at lower ${\bf k}$ modes, corresponding to the global emergence of oscillons, prolongs this 
approach to equipartition. The inset of Fig. \ref{S} shows the large variations in $\Pi(t)$ (dotted 
line) that arise due to the synchronous oscillations in the kinetic energy of these configurations. 
Also shown (solid line) is the average between successive peaks of $\Pi(t)$, with a plateau at 
$20 \lesssim t \lesssim 70$ that coincides with the maximum oscillon presence in the system. Thus, 
oscillon configurations serve as early bottlenecks to equipartition, temporarily suppressing the 
diffusion of energy from low ($0<|{\bf k}|\lesssim 0.8$) to higher modes.

We have investigated the nonequilibrium evolution of a scalar field with a double-well potential. For a
range of temperatures, the approach to equilibrium is characterized by three stages: first, the
synchronized emergence of oscillons; second, the loss of the initial synchronicity, but the persistence 
of oscillons; and third, their disappearance as the system approaches equipartition. It would be 
interesting to investigate the possibility of controlling the 
duration of the synchronization stage and search for this emergent behavior in laboratory systems, 
ranging from vibrating grains to ferromagnetic materials. 

MG was supported in part by NSF grants PHY-0070554 and PHY-0099543.

\end{document}